\documentclass[reprint, aps,prd,twocolumn,showpacs,showkeys, 10pt, longbibliography, nolinenumbers, floatfix]{revtex4-1}

\usepackage{amsmath}
\usepackage{graphicx}
\usepackage{float}
\usepackage{dcolumn}
\usepackage{multirow}
\usepackage{natbib}
\usepackage{changes}
\usepackage{comment}
\usepackage[colorlinks = true,linkcolor = blue,urlcolor  = blue,citecolor = blue,anchorcolor = blue]{hyperref}
\usepackage{enumerate}

\begin{document}

\title{ Neutron Dark Decay in Neutron Stars: The Role of the Symmetry Energy}

\author{M. Divaris}

\author{Ch.C. Moustakidis}


\affiliation{Department of Theoretical Physics, Aristotle University of Thessaloniki, 54124 Thessaloniki, Greece  }

\begin{abstract}

We conduct a systematic investigation of the influence of the nuclear symmetry energy on the proposed neutron decay into dark matter particles  within the cores of neutron stars. Unlike the majority of  previous studies that considered only pure neutron matter, the present analysis is extended to encompass $\beta$-stable nuclear matter. Furthermore, in relation to previous studies, the interactions between dark matter and baryons are incorporated and systematically studied regarding their effect on the structure of neutron stars. Our findings indicate that the nuclear symmetry energy plays a critical role in shaping the total equation of state (EoS) for dense neutron star matter containing dark sector components. The strength of interactions among dark matter particles, as well as between dark matter and baryons, is shown to be pivotal in determining both the composition and the macroscopic properties of neutron stars.  The concurrent tuning of interaction strengths alongside the symmetry energy parameters may facilitate a more accurate reproduction of recent observational data relevant to neutron star properties. In any case, the extent to which the proposed dark decay of the neutron is affected by the extreme conditions prevailing in the interior of neutron stars remains an open problem.


\keywords{Neutron dark decay, Neutron stars, Dark matter, Nuclear Symmetry Energy}
\end{abstract}

\maketitle

\section{Introduction}
One of the most long-standing problems in nuclear
physics is the discrepancy between the neutron lifetime
measured in beam and bottle experiments.  In the bottle method,
neutrons are trapped, making it sensitive to the total neutron width, yielding a slightly
shorter lifetime of $\tau_n^{\rm bottle}=
879.6\pm 0.6$ s \cite{Pichlmaier-2010,Serebrov-2005,Steyerl-2012}. In contrast, the beam method measures
the number of protons produced through $\beta$-decay in a fixed volume from a neutron beam,
resulting in a longer lifetime of $\tau_n^{\rm beam}=
888.0\pm 2.0$ s \cite{Arzumanov-2015,Byrne-1990,Yue-2013}. This discrepancy  led  Fornal and  Grinstein to propose the scenario of the existence of neutron dark  decay~\cite{Fornal-2018}. In particular, they proposed as an explanation for this anomaly, a dark decay channel for the neutron, involving one or more dark matter particles in the final state (for more details see also Ref.~\cite{Fornal-2023}).

Following the hypothesis of   Fornal and  Grinstein~\cite{Fornal-2018},   in a series of papers, the effect of the   neutron dark  decay on the properties of neutron stars and finite nuclei was studied~\cite{Baym-2018,Gil-2024,Motta-2018a,Motta-2018b,Husain-2022a,Grinstein-2019,McKeen-2018,Husain-2023,Husain-2025,Shirke-2023,Das-Burgio-2025,Tan-2019,Cline-2018,Fornal-2020,Darini-2023,Strumia-2022,Ejiri-2019,Ivanov-2019,Veselsky-2025} (see also the reviews~\cite{Zhou-2023,Gardner-2023,Bramante-2024,Grippa-2025}). A common conclusion of the above studies is that dark matter must exhibit sufficiently repulsive interactions to overcome the softening of the equation of state due to adding a
second species, thus accommodating $2 M_{\odot}$ neutron stars.
According to Ref.~\cite{Baym-2018} such a scenario makes very specific demands on the dark
matter self-interaction strength as a function of the dark
matter fermion density. Moreover, in Ref.~\cite{Grinstein-2019} the authors demonstrated   that appropriate dark matter-baryon interactions can
accommodate neutron stars with mass above two solar masses. In the same direction lies the finding of Ref.~\cite{McKeen-2018} according to which the existence of neutron stars with a mass greater than  $0.7 M_{\odot}$ places severe
constraints on such particles, requiring them to be heavier than 1.2 GeV or to have strongly repulsive
self-interactions.

In the present work we consider the decay channel  
$
    n\longrightarrow  \chi +\phi
$
where $\chi$ is the dark matter (DM) particle and  $\phi$ is a dark boson that has a very small mass.  It is worth mentioning here that other additional mechanisms of decay have been proposed. These mechanisms are $n\longrightarrow  \chi +e^+ +e^-$ and $n\longrightarrow  \chi +\gamma$. However, the above mechanisms have been rejected after experimental measurements~\cite{Tang-2018,Sun-2018}.

It is important to emphasize that the mechanisms described above have thus far only been studied under laboratory conditions. In contrast, the extreme environment within neutron stars, characterized by vastly different pressures, densities, temperatures, and surrounding matter could significantly alter the feasibility of neutron dark decay. These conditions might either enhance or suppress the process entirely. As such, this represents an additional and critical parameter in the problem, one that cannot be ignored, yet currently cannot be resolved with scientific certainty.

There are many reasons to believe that if this  neutron dark  decay is taking place then it is also involved in the properties of neutron stars. 
Due to $\beta$-equilibrium, neutrons, protons, electrons and dark particles will coexist inside the neutron star with their proportions determined by the respective chemical reactions and equilibria. In this case, we expect the nuclear symmetry energy (NSE) that determines the proton-neutron ratio in neutron star matter to play an important role. We also expect it to play a role in the ratio of dark particles and thus in the form of the equations of state involving dark matter  and the form of the corresponding mass–radius configurations.

The NSE  is one of the most fundamental quantities relevant to the study of both neutron-rich finite nuclei and neutron stars (for a comprehensive review see Refs.~\cite{Baldo-2016,Steiner-2005,Sammarruca-2013,Lattimer-2023,Lattimer-2014,BaoLI-2014a,Horowitz-2014,Bao-2021,Bao-2022,Burgio-2021}). This quantity is directly related to the isovector character of the nuclear forces and  exhibits a strong dependence on the baryonic density. 
Both theoretical
and experimental efforts are focused on the study of a possible
correlation of the two main parameters concerning the NSE, that is the slope parameter   $L$ and its value at the saturation density of nuclear matter $J$, with various nuclear properties~\cite{Baldo-2016,Steiner-2005,Sammarruca-2013,Lattimer-2023,Lattimer-2014,BaoLI-2014a,Horowitz-2014,Bao-2021,Bao-2022,Burgio-2021}. These properties include mainly nuclear masses,  the neutron skin thickness, the nuclear dipole polarizability, the giant and pygmy dipole resonance  energies, flows in heavy-ion collisions and isobaric analog states (for a comprehensive analysis see Ref.~\cite{Lattimer-2023}). Moreover, there is  a variety of neutron star properties
that are sensitive to NSE   e.g. the radius and the maximum mass, the crust-core transition density and consequently  the  crust's  thickness, the thermal relaxation time, the various neutrino processes related with the cooling, and reaction rates involved in the astrophysical r-process~\cite{Lattimer-2023}. In the present study, inspired by the previous study developed in Refs.~\cite{Sotani-2014,Sotani-2022,Divaris-2024},  we  parameterize the equation of state (EoS) which describes the asymmetric and symmetric nuclear matter with the help of the parameter $\eta=(K_0 L^2)^{1/3}$, where $K_0$ is the incompressibility and $L$ the slope parameter. The parameter $\eta$  is a regulator of the stiffness of the equation of state.

The main motivation of this work is to study in a systematic way the effect of the NSE  on the  equations of state of dense nuclear matter involving neutron dark decay as well as on the bulk properties of the corresponding neutron stars. In this effort we consider that the produced dark particles, which are in equilibrium within the nuclear matter, constitute a gas of repulsively interacting fermions.
Moreover, in order to enrich our study we consider the case of dark matter-baryon interaction (see also Ref.~\cite{Grinstein-2019}). 
In each case, the effect of the three main parameters is studied, that is, the one associated with the symmetry energy and the other two with the strength of the repulsive self-iteraction and the repulsive dark matter-baryon interaction. To the best of our knowledge, the influence of symmetry energy on neutron dark decay has not been systematically investigated, particularly with regard to all potential interactions between dark matter and both itself and baryonic matter. The results are useful not only because they give an estimate of the effects of dark matter on the properties of NSs, but also because one can place constraints on the strength of these interactions with the help of observations.

The paper is organized as follows: In Section II, we briefly present the concept of  neutron dark  decay in neutron stars. Section III is dedicated to the construction of the equation of state, while Section IV presents and discusses the results. Finally, Section V concludes the study with our final remarks.

\section{Neutron dark  decay}
The dominant neutron decay channel is the classical $\beta$-decay where
\begin{equation}
n\longrightarrow p+e^{-}+\bar{\nu}_e
\label{beta-1}
\end{equation}
In the present work, following the suggestion in Ref.~\cite{Fornal-2018}, we employ an addition mechanism, concerning the neutron decay according to (see also Refs.~\cite{Gil-2024,Motta-2018a,Motta-2018b,Shirke-2023})
\begin{equation}
n\longrightarrow \chi+\phi
\label{beta-2}
\end{equation}
where $\chi$ is a dark half-spin fermion with baryon number 1, and $\phi$ is a light dark boson that escapes from the neutron star interior and  therefore does not participate in the construction of the equation of state of  the star. Following the analysis of Ref.~\cite{Fornal-2023} there are some restrictions in the range of dark particle masses. So, nuclear stability demands 
\begin{equation}
  937.993 \ {\rm MeV} < m_{\chi}+m_{\phi} < 939.565 \ {\rm MeV}
  \label{mass-range}
\end{equation}
The light dark particle $\phi$ is considered to escape from neutron stars. In order to allow for the neutron decay channel into $\phi_\chi$ while ensuring the stability of nuclei, one must impose~\cite{Gil-2024}.
Moreover, the mass of the DM particle is bounded from below, $m_{\chi}>937.993$ MeV to prevent the decay of $^9$Be
triggered by the neutron dark decay. 
The final state fermion $\chi$ and  the scalar $\phi$ can be dark matter candidates if they  are stable, and this condition is ensured  when 
\begin{equation}
|m_{\chi}-m_{\phi}|<m_p+m_e=938.783 \ {\rm MeV}
\label{cond-2}
\end{equation}
Finally, we assume  $m_{\phi}=0$ and we fix the  particle mass at $m_{\chi}=938$ MeV.

\section{Equation of state}
\subsection{The hadronic model}

The key quantity in our calculations is  the energy per particle of asymmetric nuclear matter, where in a  good approximation, at  least for densities close to the saturation density, is given by the expression~\cite{Lattimer-2023,Lattimer-2014,Divaris-2024}
\begin{equation}
 E(n,\alpha)=E_0+\frac{K_0}{18 n_0^2}  \left(n-n_0\right)^2+ S(n)\alpha^2
 \label{En-per-bar}
\end{equation}
where $\alpha=(n_n-n_p)/n$ is the asymmetry parameter, $n$ is the total baryon density $n=n_n+n_p$ with $n_n$ and $n_p$ the number densities of neutrons and protons respectively and   $n_0$ is the  saturation density. Moreover $E_0=E(n_0,0)$ is the energy per particle at  $n_0$,  $K_0$ is the  incompressibility and $S(n)$ is the nuclear symmetry energy.  In particular,  $S(n)$ can be developed in a series around the saturation density
\begin{equation}
S(n)=J+\frac{L}{3n_0} (n-n_0)+\frac{K_{\rm sym}}{18n_0^2} (n-n_0)^2+\cdots
 \label{Sym-1}   
\end{equation}
where  $J=S(n_0)$.  The slope parameter $L$ is related to  the first derivative  and  $K_{\rm sym}$ to the second derivative of the NSE according to the definitions 
\begin{equation}
 L= 3n_0\left (\frac{dS(n)}{dn}  \right)_{n=n_0}
 \label{L-1}
\end{equation}
\begin{equation}
 K_{\rm sym}= 9n_0^2\left (\frac{d^2S(n)}{dn^2}  \right)_{n=n_0}
 \label{K0-1}
\end{equation}
In the present work, we will omit the third term in the expansion~(\ref{Sym-1}) which has a small contribution compared to the others. Now the  
 corresponding energy density ${\cal E}=nE$, which is the key quantity in the present study and essentially serves to {\it bridge} the properties of finite nuclei, nuclear matter and neutron star matter reads  
\begin{eqnarray}
 {\cal E}_b(n,\alpha) &=& (m_nc^2 +E_0 )n+\frac{K_0}{18 n_0^2}n  \left(n-n_0\right)^2 \nonumber \\
 &+&\left(J+\frac{L}{3n_0}
 (n-n_0)\right)n\alpha^2 
 \label{EnDen-per-bar}
\end{eqnarray}

In the present study $K_0$ and $L$ are parameters which combine in a single one, that is, $\eta=(K_0L^2)^{1/3}$~\cite{Sotani-2014,Sotani-2022,Divaris-2024}.
Moreover, we consider that  $n_0=0.16$ fm$^{-3}$,   $m_nc^2=939.565$, $m_pc^2=938.272$ MeV, $E_0=-16 $ MeV and $J=30$ MeV.

\subsection{The dark matter model}
Regarding the DM particles, we assume they are  fermions that interact with each other through a repulsive force.
 We consider a Yukawa-type interaction for this purpose~\cite{Nelson-2019}
\begin{equation}
V(r)=\frac{{\rm g}_{\chi}^2 (\hbar c)}{4\pi r} \exp\left[-\frac{m_{\phi}c^2}{\hbar c}r\right]
\label{Yukawa-1}
\end{equation}
where ${\rm g}_{\chi}$  and $m_{\phi}$ are   the coupling constant  and  the mediator mass respectively. The contribution on the energy density of the self-interaction is given by
\begin{equation}
{\cal E}_{SI}(n_{\chi})=\frac{(\hbar c)^3}{2z_{\chi}^2} n_{\chi}^2  
\end{equation}
where $z_{\chi}=m_{\phi}c^2/{\rm g}_{\chi}$ (in units MeV).
The total energy density of the dark matter particle ${\cal E}_{\chi}(n_{\chi})$ is given by
\begin{equation}
{\cal E}_{\chi}(n_{\chi})=m_{\chi}c^2 n_{\chi} +\frac{(\hbar c)^2(3\pi^2 n_{\chi})^{5/3}}{10\pi^2m_{\chi} c^2}+ \frac{n_{\chi}^2(\hbar c)^3}{2z_{\chi}^2}
\label{Baym-2}
\end{equation}
The  parameter $z_{\chi}$ is treated as a free parameter related with the strength of the self-interaction. For comparison with similar studies this parameter is related with the interaction strength $G_{\chi}$  by
$z_{\chi}=\hbar c/\sqrt{G_{\chi}}$~\cite{Das-Burgio-2025}.
Also, it is worth mentioning here that there exist astronomical constraints on the values of $G_{\chi}$, with typical values 4-135 fm$^{2}$ \cite{Das-65,Das-66,Das-67,Das-68,Das-69}, which correspond to the values used in the present study (for a detailed analysis see Ref.~\cite{Das-Burgio-2025}). The only exception is the value $z_{\chi}=10$ MeV which refers to a very strong interaction and can be considered as an upper limit in the present study. 

In order to enrich our study, we consider also the case of repulsive  effective
potential between the neutron (proton) and the DM particles  through the
exchange of the light scalar $\tilde{\phi}$ and  a Yukawa-type interaction for this purpose~\cite{Gil-2024}
\begin{equation}
V(r)=\frac{{\rm \tilde{g}}_{\chi} {\rm g}_{i}(\hbar c)}{4\pi r} \exp\left[-\frac{m_{\tilde\phi}c^2}{\hbar c}r\right], \quad i=n,p
\label{Yukawa-1}
\end{equation}
where ${\rm \tilde{g}}_{\chi}$  and $m_{\tilde{\phi}}$ are   the coupling constant  and  the mediator mass respectively.  The contribution on the energy density due to neutron(proton)-DM interaction is given by
\begin{equation}
{\cal E}_{\chi i}(n_{\chi})=\frac{n_{\chi}n_i(\hbar c)^3}{z_{\chi i}^2}, \quad i=n,p
\end{equation}
where $z_{\chi i}\equiv m_{\tilde{\phi}}c^2\sqrt{2}/\sqrt{|{\rm \tilde{g}}_{\chi}{\rm g}_{i}|}$. Finally, the total  baryon-dark matter interaction term is given by
\begin{equation}
{\cal E}_{\rm int}(n_n, n_p, n_{\chi})=\frac{n_{\chi}n_n(\hbar c)^3}{z_{\chi n}^2}+\frac{n_{\chi}n_p(\hbar c)^3}{z_{\chi p}^2}
\label{int}
\end{equation}
The parameter ranges for  $z_{\chi}$ and $z_{\chi i}$ are constrained and thoroughly discussed in Ref.~\cite{Gil-2024}, taking into account both the bulk properties of neutron stars (such as mass and radius) and experimental data related to neutron beta decay.

\subsection{The neutron star equation of state}
We consider neutron stars to consist mainly of neutrons and a proportion of protons and electrons with respect to $\beta$-equilibrium. Dark particles arise from the dark decay of neutrons and constitute a gas of interacting fermions (the case of interaction with baryons is also considered in the present study). 
To be more specific the two reactions below take place simultaneously
\begin{equation}
  n\longrightarrow  \chi+\phi, \qquad
   n\longrightarrow p+e^{-}+\bar{\nu}_e
   \label{both-reaction}
\end{equation}
where we consider that the boson $\phi$ and neutrino $\bar{\nu}_e$
escape form the star and do not contribute to the energy density and pressure of neutron star matter. 
The reactions (\ref{both-reaction})
in the language of chemical potentials are written 
\begin{equation}
\mu_n=\mu_{\chi}, 
\qquad \mu_n=\mu_{p}+\mu_e
\label{chem-eq-1}
\end{equation}
The above expressions with the demand of charge equilibrium
\begin{equation}
n_p=n_e  
\label{Charg-neytral-2}
\end{equation}
lead to the construction of the equation of state of neutron star-dark matter mixing.  The chemical potentials are defined as
\begin{equation}
\mu_i=\frac{\partial {\cal E}_{\rm tot}(n_n,n_p,n_{\chi},n_e)}{\partial n_i}, \quad i=n,p,\chi,e 
\label{chem-definion}
\end{equation}
where ${\cal E}_{\rm tot}(n_n,n_p,n_{\chi},n_e)$ is the total energy density which reads 
\begin{eqnarray}
{\cal E}_{\rm tot}(n_n,n_p,n_{\chi},n_e)&=&{\cal E}_{\rm b}(n_n,n_p)+{\cal E}_{\chi}(n_{\chi})
\nonumber\\
&+& {\cal E}_{\rm int}(n_n,n_p,n_{\chi}) + {\cal E}_{\rm e}(n_e)
\label{Energ-1-2}
\end{eqnarray}
where  the energy density and pressure of the electrons are given by 
\begin{equation}
{\cal E}_e(n_e)=\frac{\hbar c}{4\pi^2}\left(3\pi^2 n_e  \right)^{4/3}, \qquad P_e=\frac{{\cal E}_e}{3}
\label{Pre-ele}
\end{equation}
{\it Calculation recipe:}
We consider the neutron density $n_n$ as an independent variable.  For each value of $n_n$ we solve numerically the equations (\ref{chem-eq-1})
 with the help of Eq.~(\ref{chem-definion})
and calculate the corresponding values of $n_p$ and $n_{\chi}$ as a function of $n_n$. Then we calculate the corresponding value of  the energy  density given by Eq.~(\ref{Energ-1-2}). Finally,  the  total pressure is given by 
\begin{eqnarray}
P_{\rm tot}(n_n,n_{\chi})&=&n_p\mu_p+n_n\mu_n+n_{\chi}\mu_{\chi}+n_e\mu_e\nonumber \\
&-&{\cal E}_{\rm tot}(n_n,n_p,n_{\chi},n_e)
\label{Pres-1-1}
\end{eqnarray}
For the parameters $L$, $K_0$ and  $\eta$ we use the parametrization used in our previous work~\cite{Divaris-2024} (see also Table~\ref{Table-1}). 

\section{TOV equations and tidal deformability}
The bulk properties of neutron stars, such as the mass and  radius  are determined by solving the coupled Tolman–Oppenheimer–Volkoff (TOV) equations~\cite{Shapiro:1983du,Haensel2007NeutronS1,schaffner-bielich_2020}. As outlined in the preceding section, the equation of state, which constitutes the fundamental input of the TOV equations, is formulated by simultaneously accounting for both conventional beta decay and neutron dark decay channels.

Moreover, in recent years, valuable insights have been gained from observing gravitational waves produced by the mergers of black hole–neutron star and neutron star–neutron star binary systems. These events provide an opportunity to measure various properties of neutron stars. Notably, during the inspiral phase of binary neutron star systems, tidal effects can be detected. More specifically, the tidal Love number $k_2$ 
characterizes how a neutron star responds to an external tidal field, depending on both its mass and the applied equation of state (EoS). The exact relation governing these tidal effects is presented below~\cite{Flanagan-08,Hinderer-08}
\begin{equation}
Q_{ij}=-\frac{2}{3}k_2\frac{R^5}{G}E_{ij}\equiv- \lambda E_{ij}
\label{Love-1}
\end{equation}
where $\lambda$ is the tidal deformability. 
In addition, an important and well measured quantity by the gravitational wave detectors, which can be treated as a tool to impose constraints on the EoS, is the dimensionless tidal deformability $\Lambda$, defined as 
\begin{equation}
    \Lambda=\frac{2}{3}k_2 \left(\frac{c^2R}{GM}\right)^5=\frac{2}{3}k_2 (1.473)^{-5}\left( \frac{R}{{\rm Km}} \right)^5\left(\frac{M_{\odot}}{M}  \right)^5
\end{equation}
We notice that $\Lambda$ is sensitive to the neutron star radius, hence can provide information for the low-density part of the EoS, which is also related to the structure and properties of finite nuclei.

\section{Results and Discussion}

\begin{figure*}[ht]
  \centering
  \includegraphics[width=0.42\textwidth]{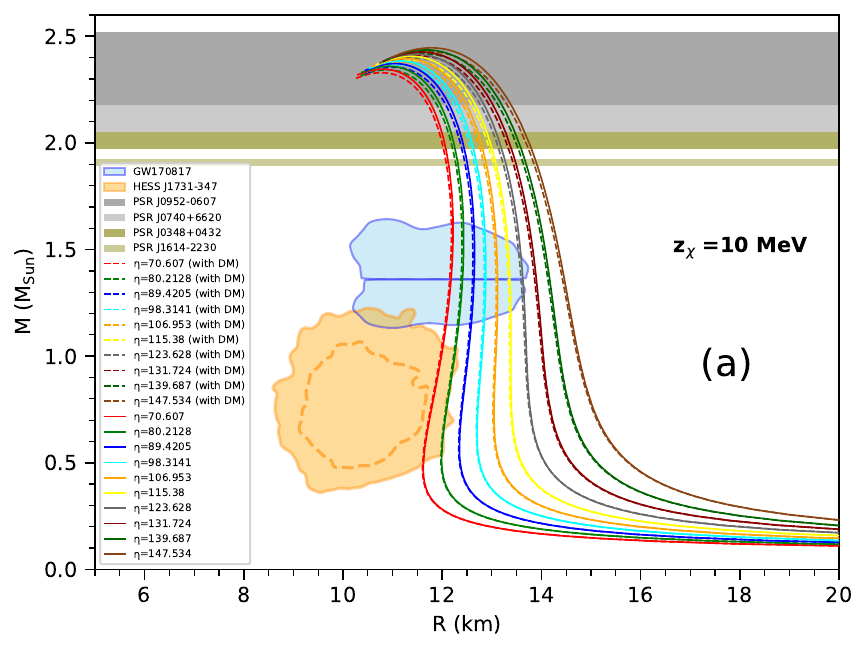}
  \hfill \includegraphics[width=0.42\textwidth]{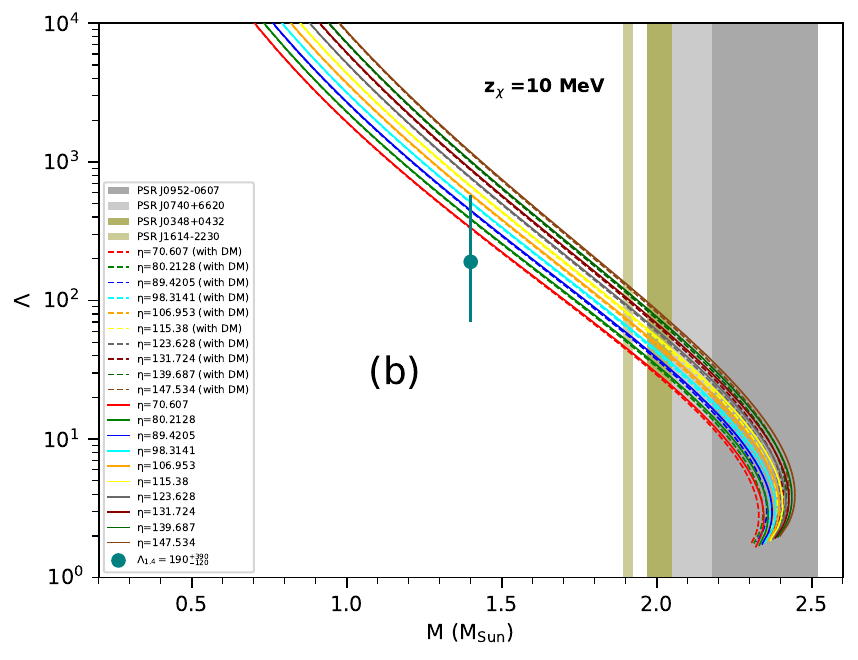}
  \includegraphics[width=0.42\textwidth]{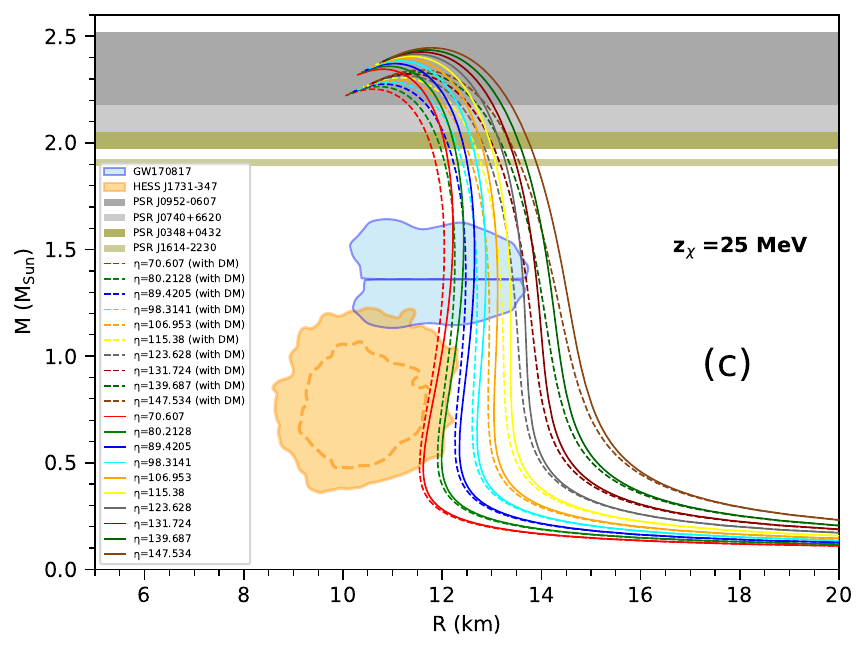}
  \hfill
  \includegraphics[width=0.42\textwidth]{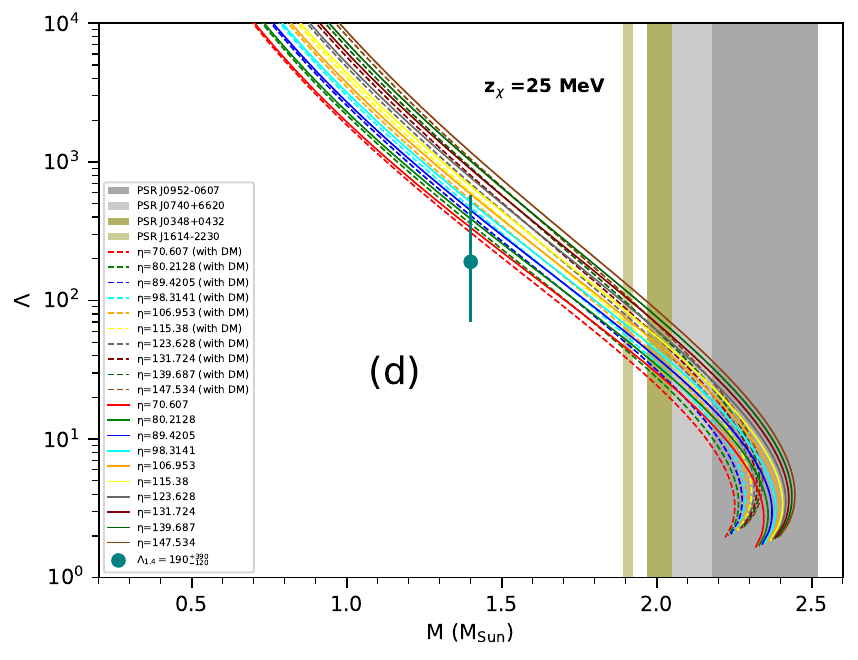}
\caption{The M-R diagrams  (left) and the tidal defomability $\Lambda$ as a function of the mass M  (right) with interaction parameter $z_{\chi}=10$ MeV (figures (a) and (b)) and $z_{\chi}=25$ (figures (c) and (d))  respectively and for various cases of the parameter $\eta$ with and without dark matter. On the left figures the shaded
regions from bottom to top represent the HESS J1731-347 remnant
\cite{Doroshenko-2022}, the GW170817 event \cite{Abbott-2019-X}, PSR J1614-2230 \cite{Arzoumanian-2018}, PSR
J0348+0432 \cite{Antoniadis-2013}, PSR J0740+6620 \cite{Cromartie-2020}, and PSR J0952-0607  
\cite{Romani-2022} pulsar observations for the possible maximum mass. On the right figures the  blue point with its corresponding
error-bar indicates the estimated value of $\Lambda_{1.4}$ (which corresponds to the mass $M=1.4 \ M_{\odot}$), provided
by the GW170817 detection \cite{Abbott-2019-X}. }
  \label{M-R-L-z10}
\end{figure*}

In Fig.~\ref{M-R-L-z10} we present the effects of dark neutron decay on the M-R diagram for various values of the parameter $\eta$ with and without the presence of the DM particle. This case corresponds to the one with the strongest interaction between dark matter particles $z_{\chi}=10$ MeV. Initially, we observe that increasing the parameter $\eta$
results in a stiffer equation of state, manifested as an increase in the maximum mass and the corresponding radius at intermediate masses, as expected. 
Clearly, in each  case, the effects of dark neutron decay are minimal. This is easily explained by the fact that, due to the strong self-interactions among dark particles, they only appear at high baryonic densities, and therefore do not significantly alter the nuclear matter equation of state. Accordingly, in Fig.~\ref{M-R-L-z10} we present the dependence of the tidal deformability $\Lambda$ on mass. Similar to the mass–radius relation, the influence of dark matter remains negligible, with the effects primarily driven by the symmetry energy.

\begin{table}[h]
\caption{The incompressibility $K_0$ (in MeV), the slope parameter $L$ (in MeV) and the corresponding values of  the parameter $\eta$ (in MeV) used in the present study. }
\begin{tabular}{ c c c cccccccc }
\hline
$ K_0 $	& 220& 224 &228& 232&  236& 240 & 244& 248&  252&  256  \\
\hline
$L$ & 40 &48 & 56&  64&  72& 80& 88&  96&  104&  112   \\
\hline
$\eta$ & 70.6 &80.2& 89.4& 98.3& 107.0&115.4& 123.6& 131.7& 139.7& 147.5 \\
\hline
\end{tabular}
\label{Table-1}
\end{table}

\begin{figure*}[ht]
  \centering
  \includegraphics[width=0.42\textwidth]{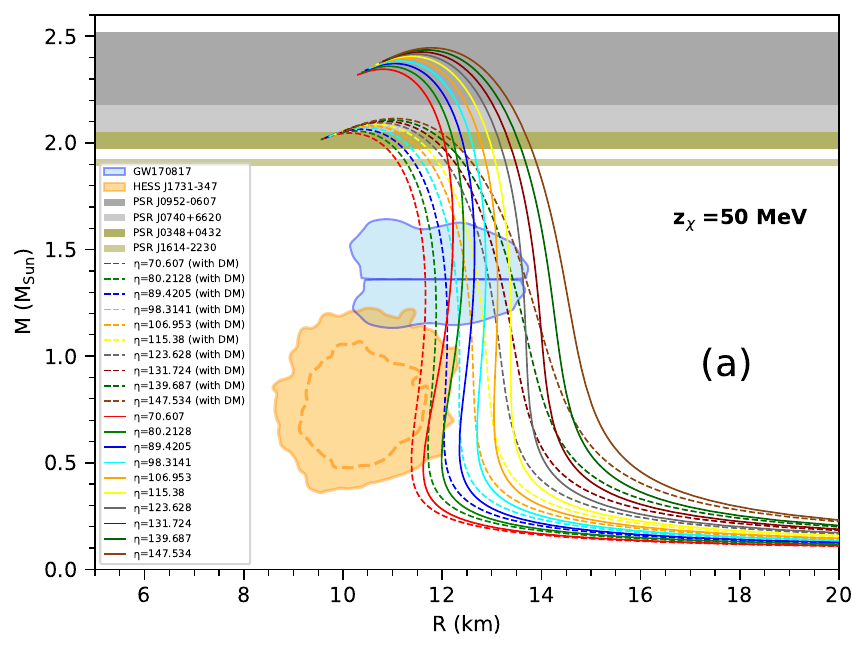}
  \hfill
  \includegraphics[width=0.42\textwidth]{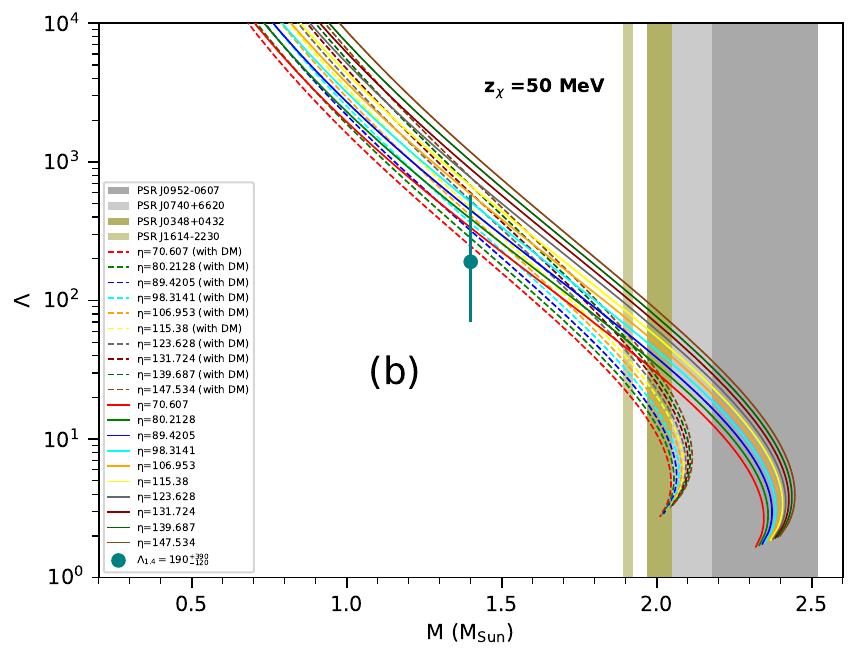}
 \includegraphics[width=0.42\textwidth]{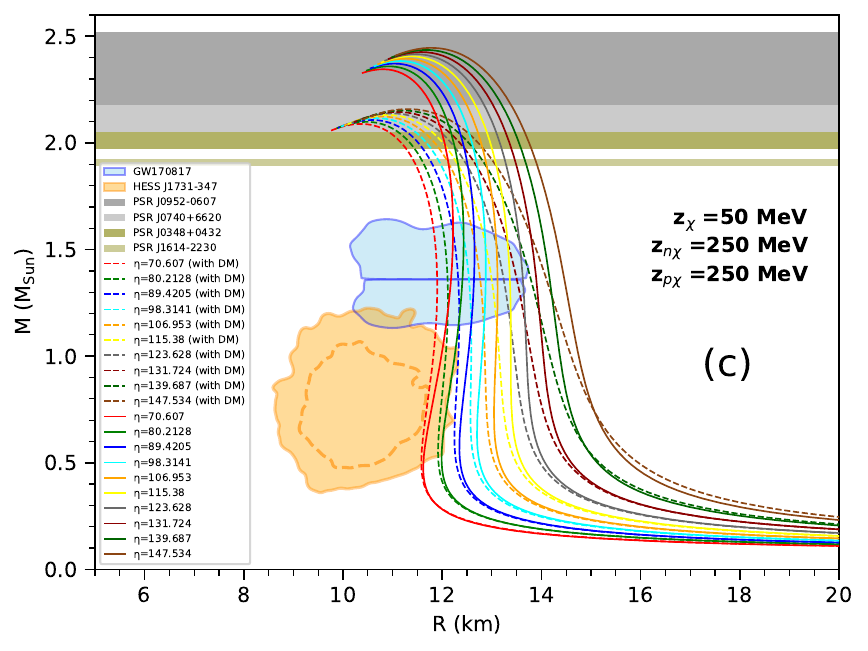}
  \hfill
  \includegraphics[width=0.42\textwidth]{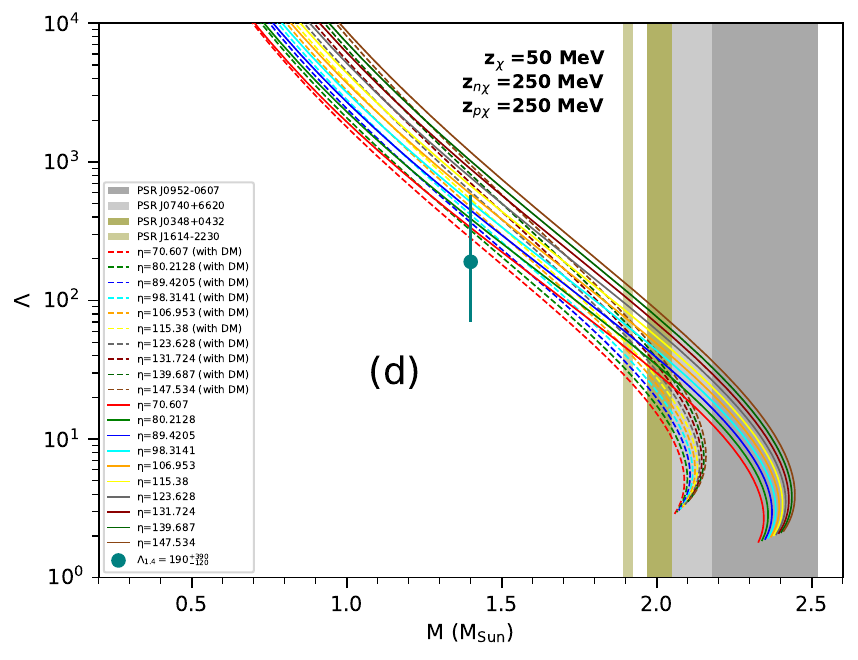}
 \includegraphics[width=0.42\textwidth]{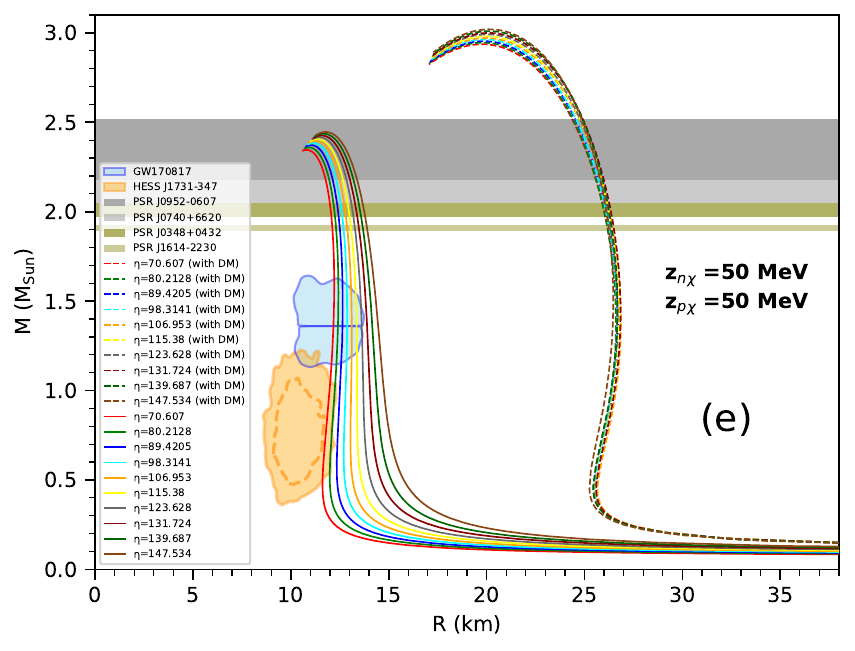}
 \hfill
  \includegraphics[width=0.42\textwidth]{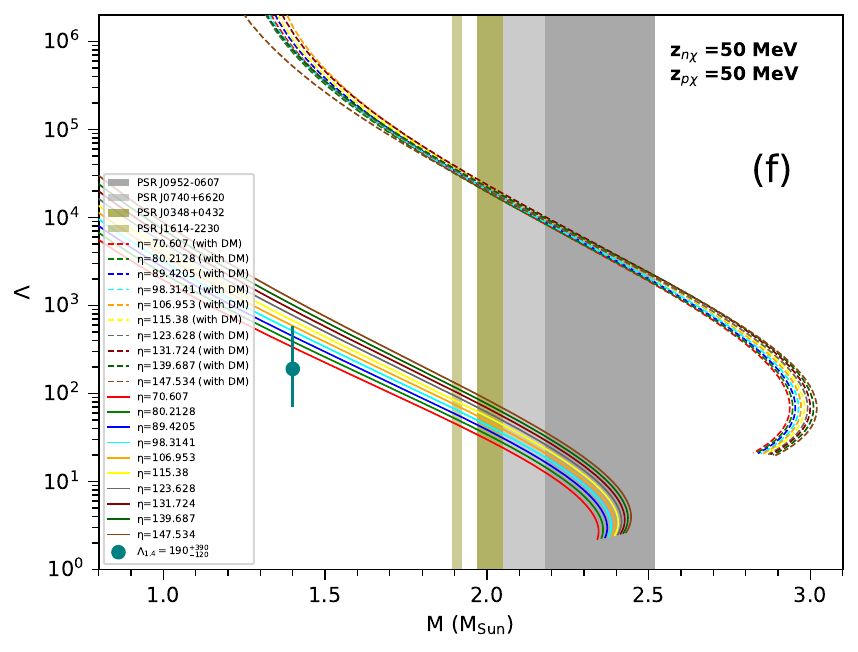}
  \includegraphics[width=0.42\textwidth]{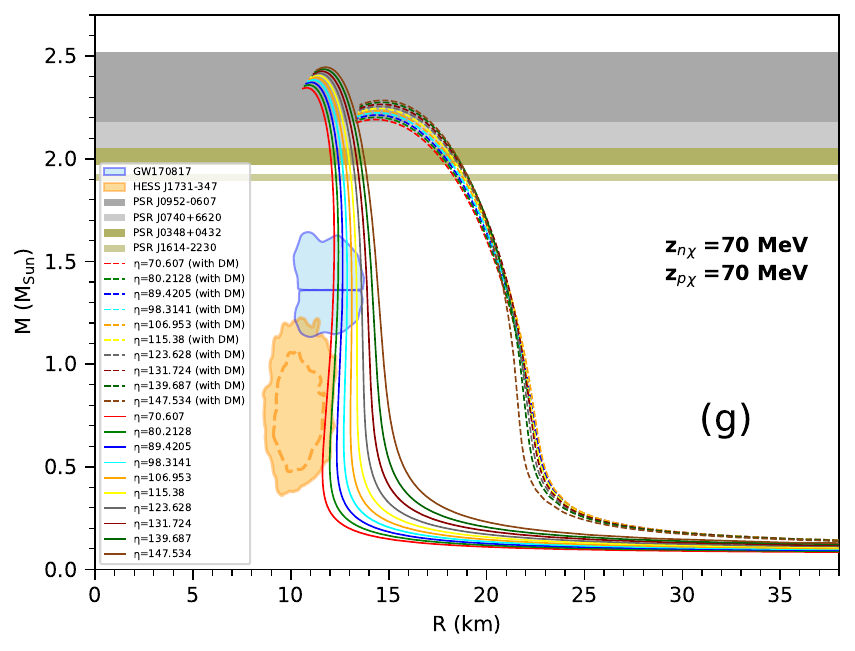}
 \hfill
  \includegraphics[width=0.42\textwidth]{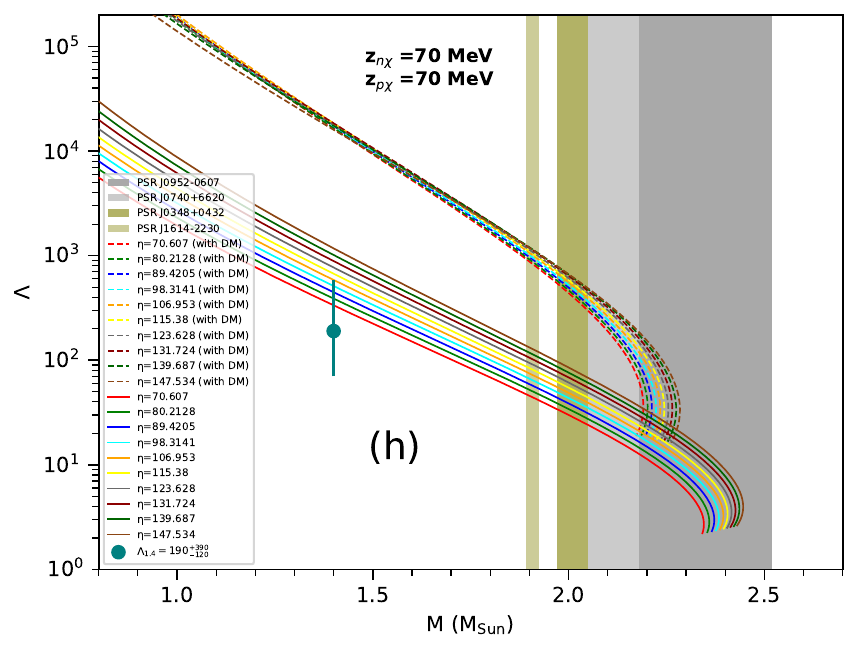}
\caption{The same with Fig.~\ref{M-R-L-z10} for four different cases of interaction that is  only self interaction with $z_{\chi}=50$ MeV (figures (a) and (b)),  self-interaction+baryon-dark matter interaction (with $z_{\chi}=50$ MeV and $z_{i\chi}=250$ MeV) (figures (c) and (d))  and  only baryon-dark matter interaction (case with $z_{i\chi}=50$ MeV (figures (e) and (f))  and case with $z_{i\chi}=70$ MeV (figures (g) and (h))) for various values of the parameter $\eta$ with and without dark matter.  }
  \label{M-R-L-z50}
\end{figure*}

\begin{figure*}[ht]
  \centering
  \includegraphics[width=0.42\textwidth]{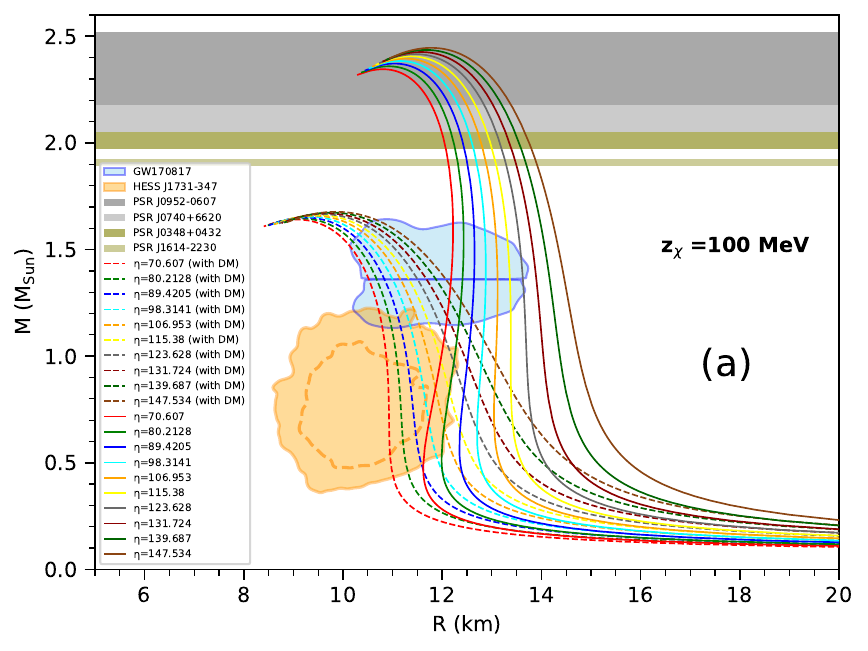}
  \hfill
  \includegraphics[width=0.42\textwidth]{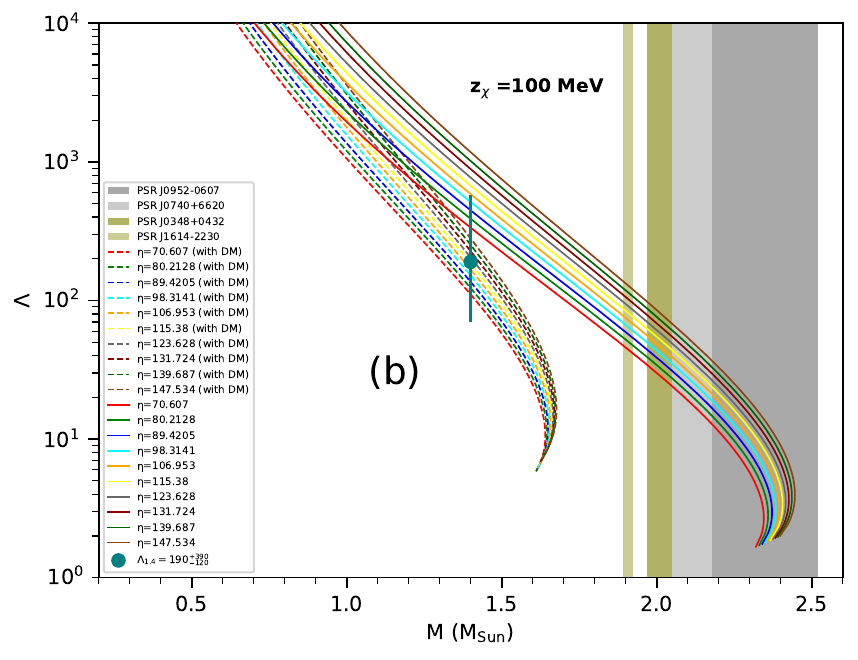}
 \includegraphics[width=0.42\textwidth]{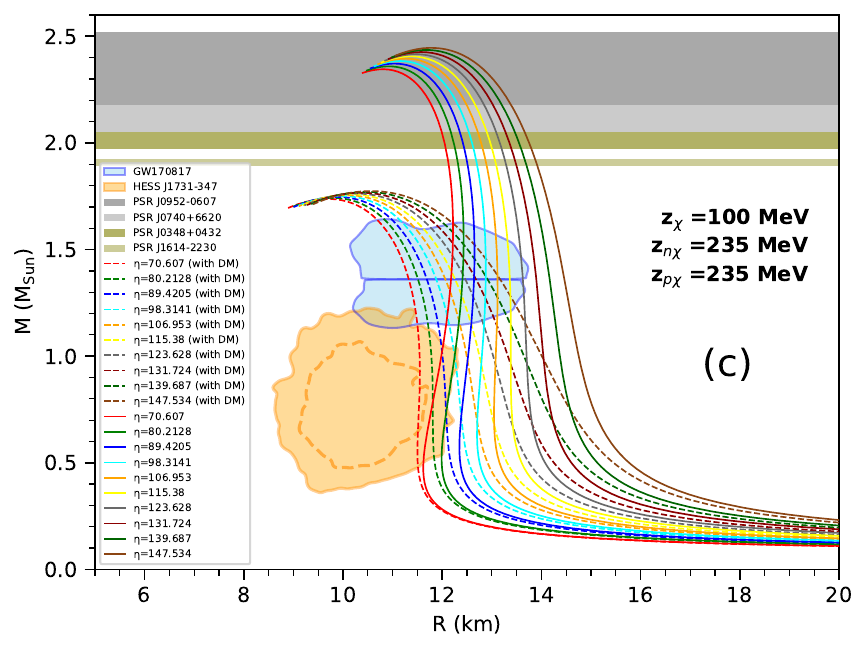}
  \hfill
  \includegraphics[width=0.42\textwidth]{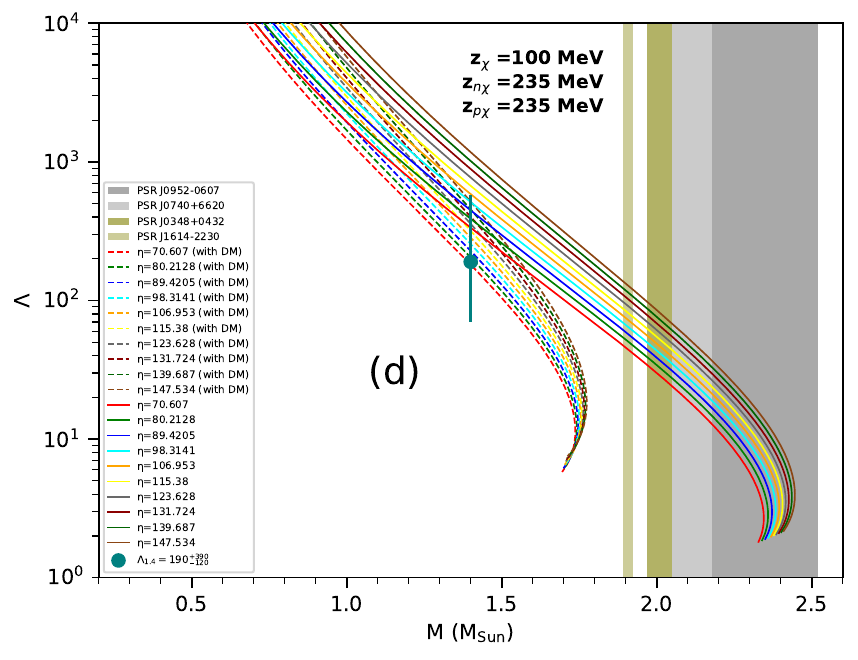} 
  \caption{The same with Fig.~\ref{M-R-L-z50} but  only self interaction with $z_{\chi}=100$ MeV (figures (a) and (b)) and   self-interaction+baryon-dark matter interaction with $z_{\chi}=100$ MeV and $z_{i\chi}=235$ MeV (figures (c) and (d)).}
  \label{M-R-L-z100}
\end{figure*}

\begin{figure*}[ht]
  \centering
  \includegraphics[width=0.3\textwidth]{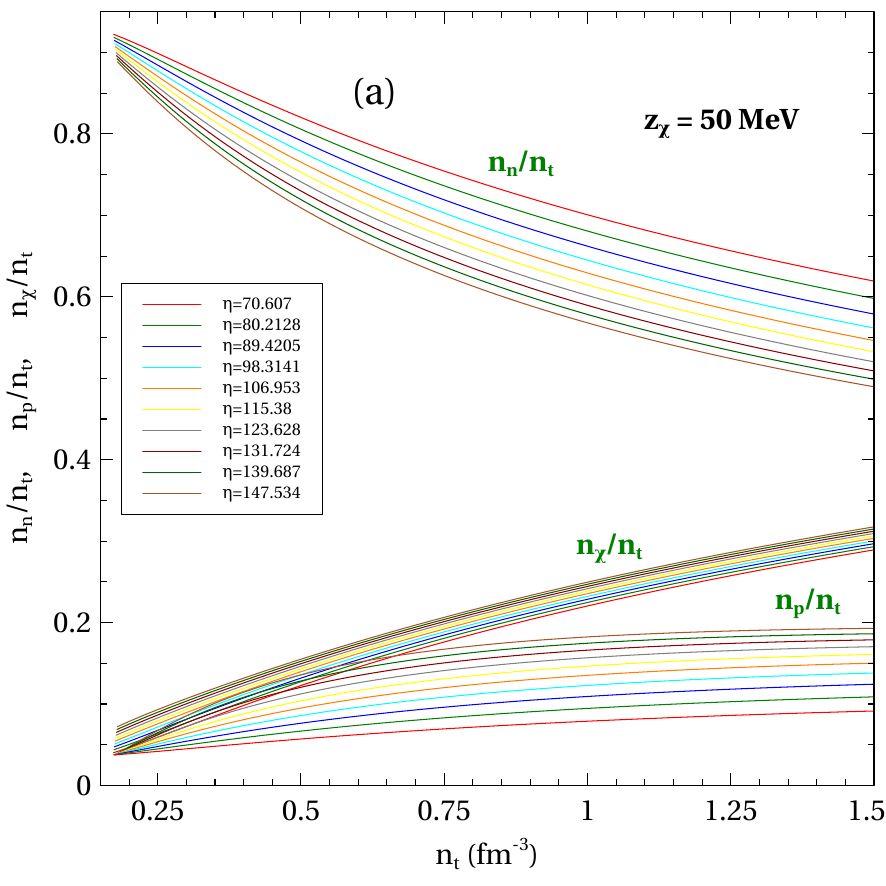}
  \includegraphics[width=0.3\textwidth]{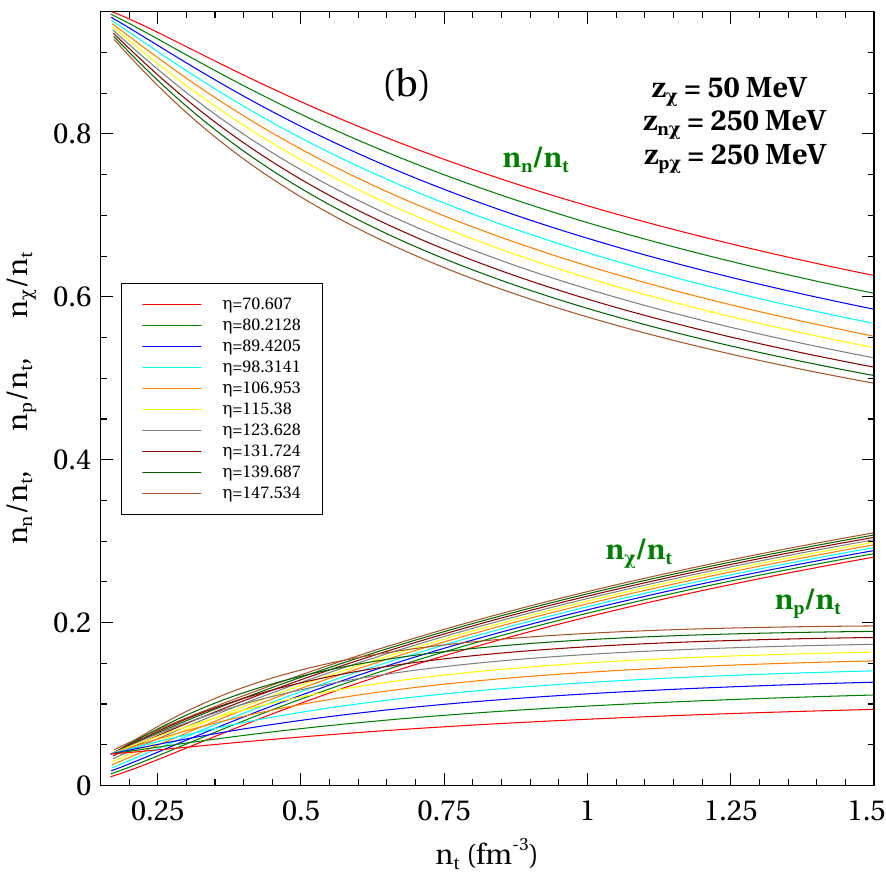}
  \includegraphics[width=0.3\textwidth]{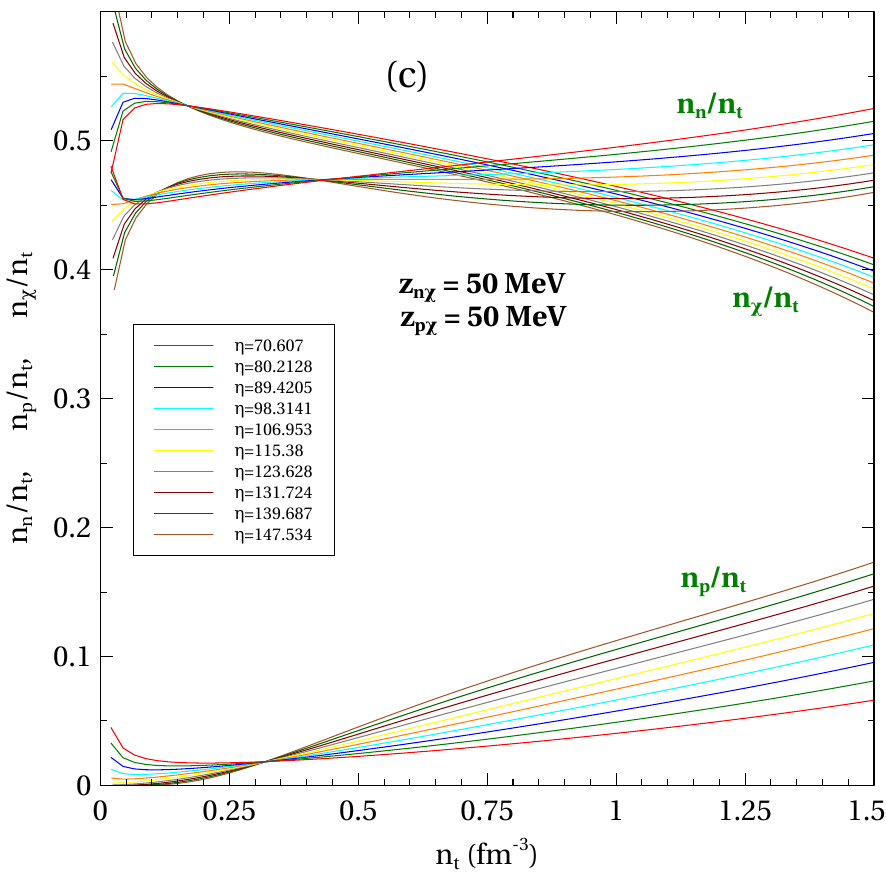}
\includegraphics[width=0.3\textwidth]{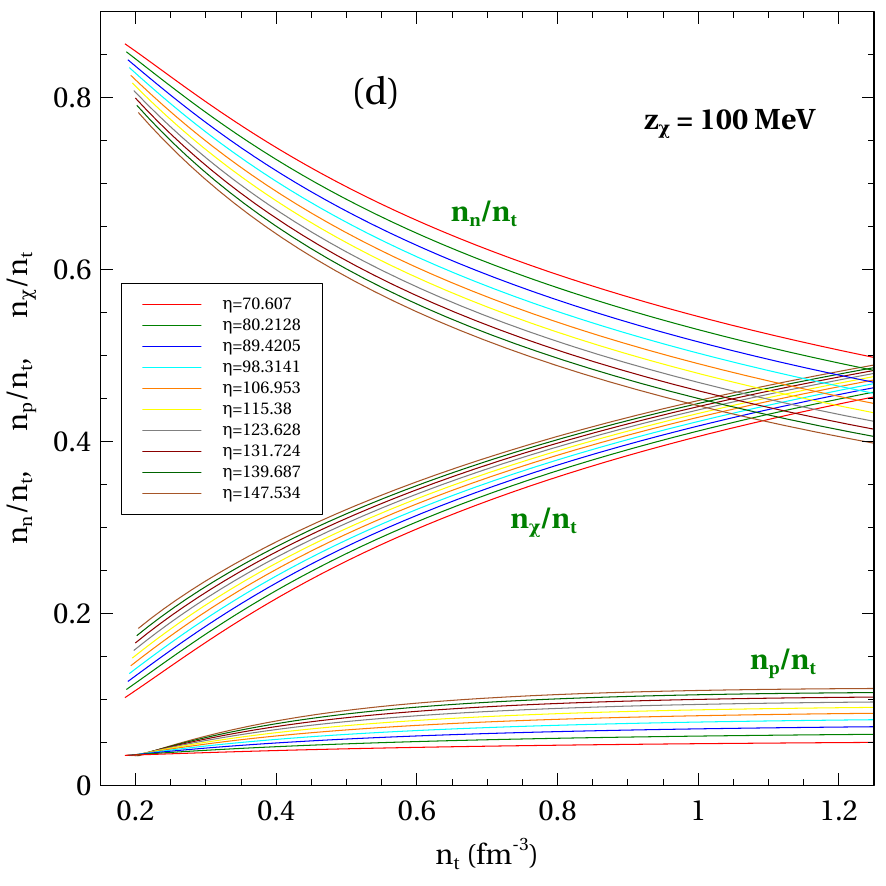}
\includegraphics[width=0.3\textwidth]{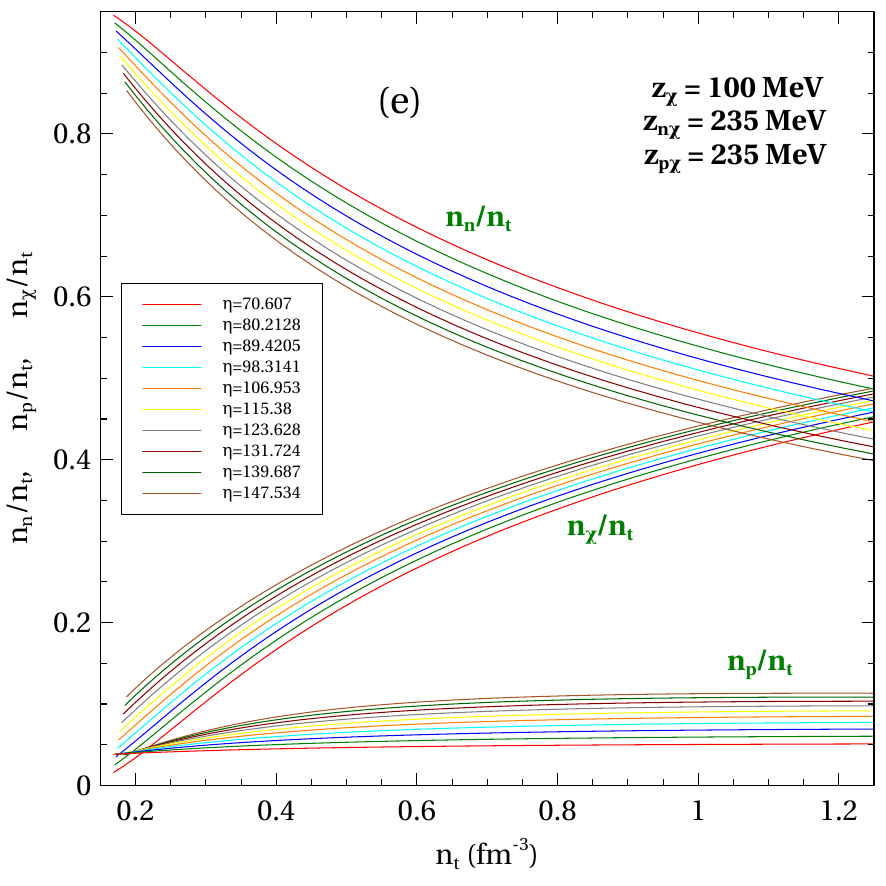}
\includegraphics[width=0.3\textwidth]{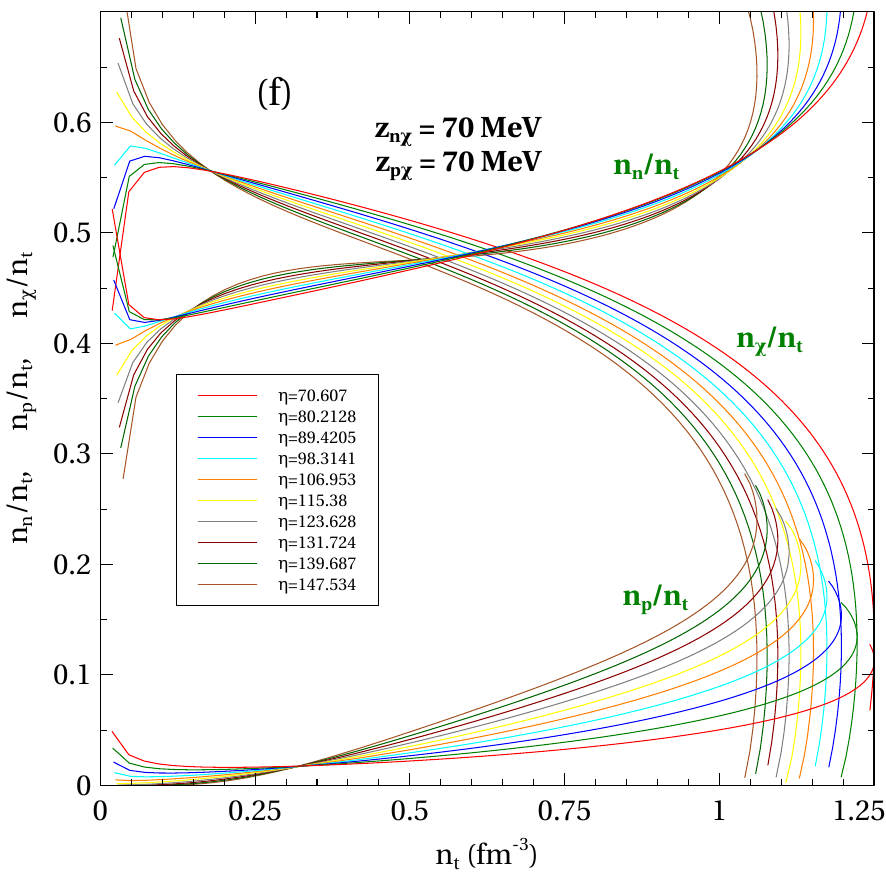}
\caption{The fractions of neutrons, protons and dark matter particles  as a function of the total number density $n_t$ and  for  various values of the interaction parameters $z_{\chi}$, $z_{n\chi}$ and  $z_{p\chi}$. }
  \label{frac-z50-z250-xn50}
\end{figure*}

\begin{figure*}[ht]
  \centering
  \includegraphics[width=0.35\textwidth]{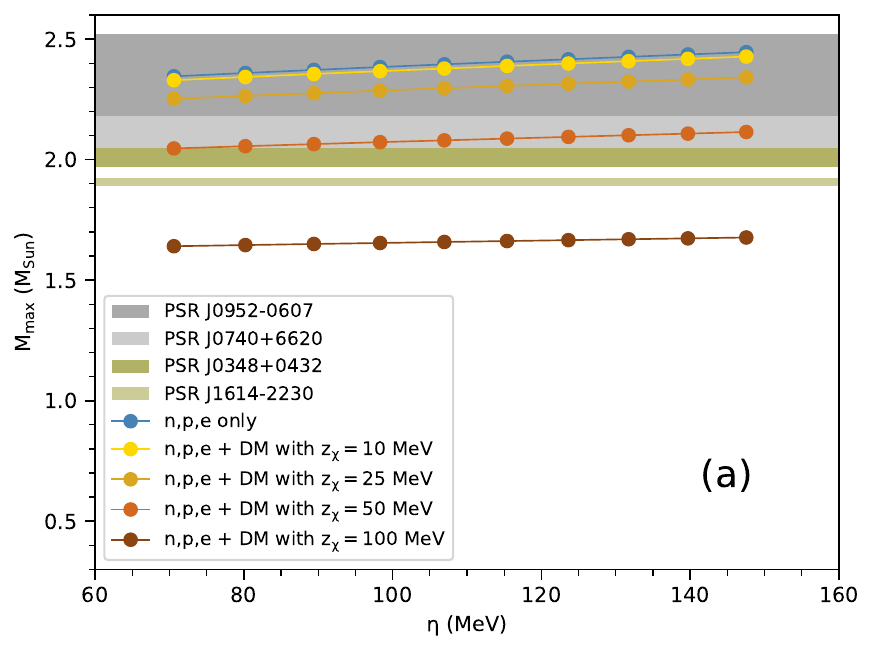}
\includegraphics[width=0.35\textwidth]{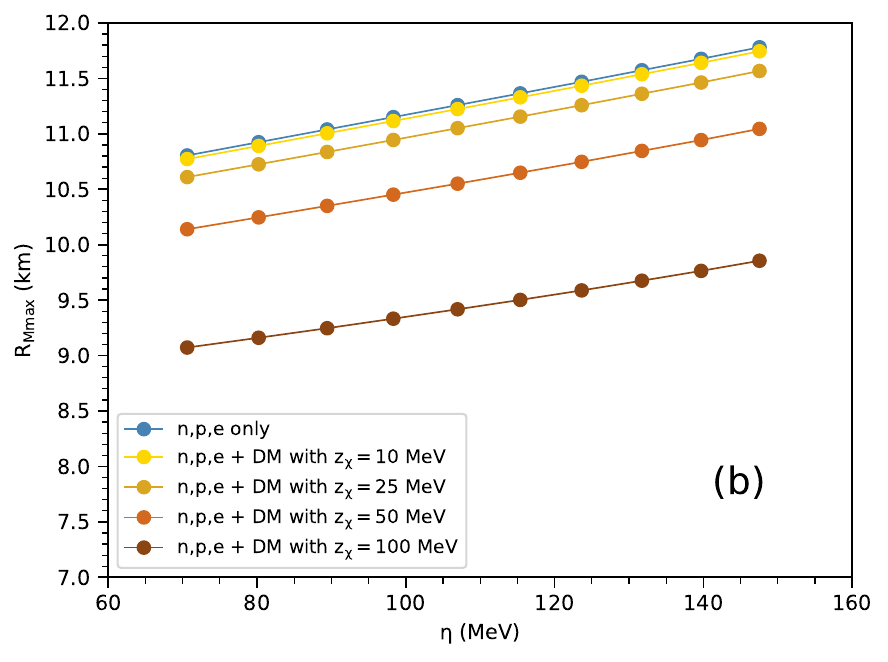}
\\ \includegraphics[width=0.35\textwidth]{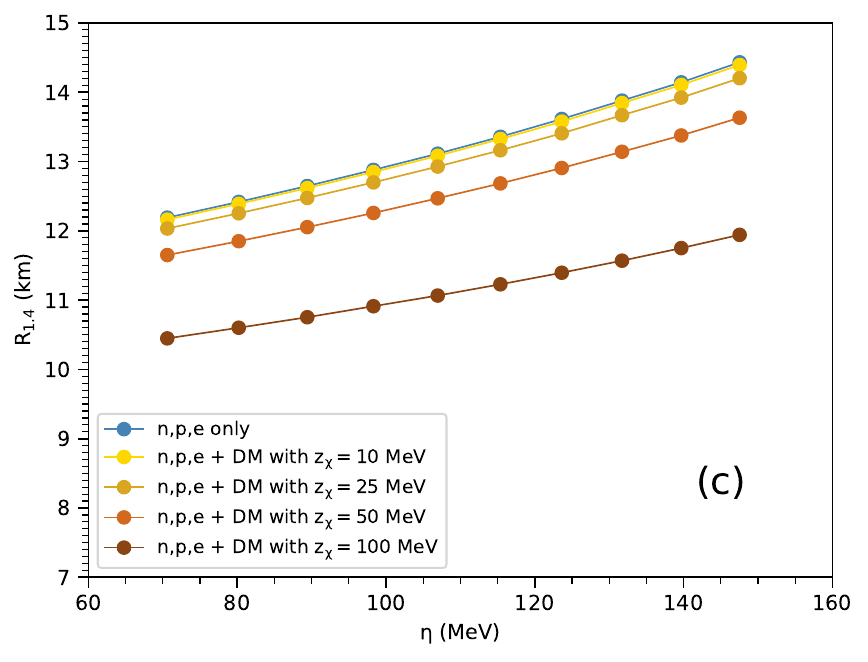}
 \includegraphics[width=0.35\textwidth]{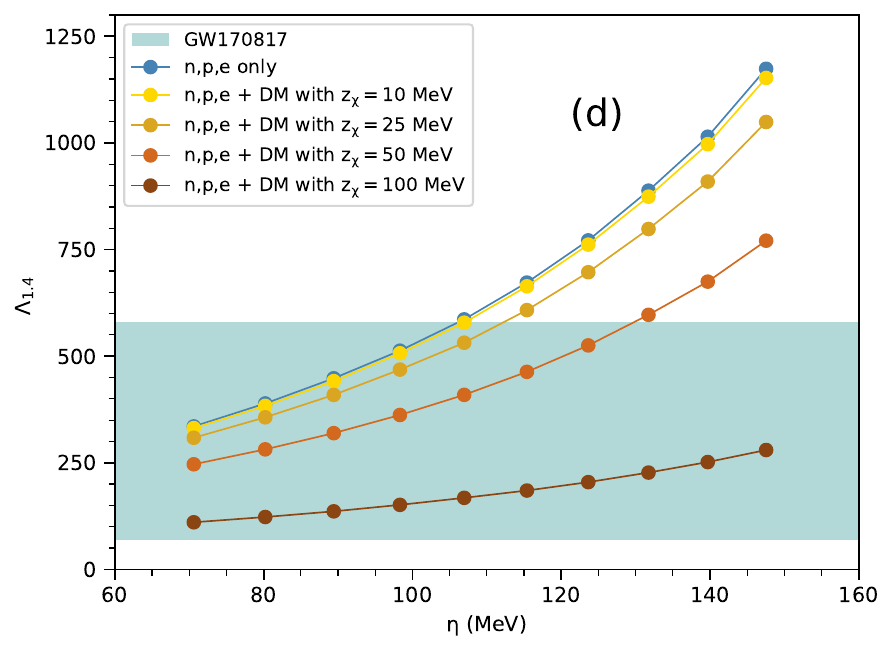}
\caption{The values of the maximum mass $M_{\rm max}$, the corresponding radius  $R_{\rm max}$, the radius  $R_{ 1.4}$, which corresponds to the mass $M=1.4 \ M_{\odot}$ and the tidal deformability   $\Lambda_{ 1.4}$ (the shaded area indicates the estimated value of $\Lambda_{1.4}$, provided
by the GW170817 detection \cite{Abbott-2019-X})  as a function of the parameter $\eta$ and for various value  of the self-interaction parameter $z_{\chi}$.}
   \label{M-R-eta-all}
\end{figure*}


The impact of dark matter on the equation of state becomes noticeable at higher values of the parameter  $z_{\chi}$ (indicating weaker interactions), as also  illustrated in Fig.~\ref{M-R-L-z10}. In this case, the equation of state softens, resulting in lower maximum masses and smaller radii in the corresponding mass–radius diagram. Interestingly, in the mass range around $\sim $1.4 $M_{\odot}$, which is most commonly observed, the spread in radii due to symmetry energy uncertainties is larger in the absence of dark matter than in its presence. In other words, the presence of dark matter acts to constrain the uncertainty in the predicted radius. This effect becomes more pronounced as the interaction strength decreases (i.e., for larger values of $z_{\chi}$ as illustrated in Figs.~\ref{M-R-L-z50} and \ref{M-R-L-z100}).

To enhance our study, we also consider the case where baryon-dark matter repulsive interactions occur alongside self-interactions. In particular, in Fig.~\ref{M-R-L-z50} we display the case where    $z_{\chi}=50$ MeV and  $z_{n\chi}=z_{p\chi}=250$ MeV. It is evident that the incorporation  of the repulsive baryon–dark matter interaction results in the corresponding equation of state becoming stiffer, leading to correspondingly larger maximum masses and radii. We reach the same results if we use a different set of interaction  parameters (see also  Fig.~\ref{M-R-L-z100}). 

We further examine the interesting case where only the interaction between dark matter particles and baryons is taken into account. The results are presented in  Figs.~\ref{M-R-L-z50}(e)-\ref{M-R-L-z50}(h). In this case, it is particularly noteworthy that the resulting equations of state become especially stiff, leading to large masses (within the mass-gap region) and correspondingly large radii. 
This constitutes a significant result that, to the best of our knowledge, has not been systematically explored to date. It effectively opens up the possibility for the existence of compact objects with masses in the mass-gap region, provided that a strong repulsive interaction between baryons and dark matter particles is taken into account (see also Ref.~\cite{Gil-2024} for a detailed discussion of the interaction range).
The present findings are in good agreement with recent related studies, although key differences exist~\cite{Vikiaris-2024,Vikiaris-2025}:
(a) the assumed origin of the dark matter component differs, and
(b) the stable configurations were obtained through the implementation of a two-fluid model.
Notwithstanding these differences, the convergence of results across independent approaches is noteworthy and strengthens the robustness of the underlying physical scenario.

It is also interesting to observe the effect of the symmetry energy, for various values of the interaction parameters, on the fractions of the fundamental particles involved in the specific composition of neutron star matter. Thus in Fig.~\ref{frac-z50-z250-xn50} we display the neutron, proton and dark particle fractions for specific values of the interaction parameters and of the parameter $\eta$ as a function of the total number density $n_t=n_n+n_p+n_{\chi}$. The most striking feature is that, while the composition of matter does not change significantly when the interaction between baryons and dark particles is taken into account, it changes dramatically when only the interaction between baryons and dark particles is considered (see Fig.\ref{frac-z50-z250-xn50}(c)). In fact, dark particles appear to dominate at lower densities, while neutrons prevail at higher densities.
In any case, the effect of the symmetry energy is evident.
Clearly, this composition of matter is also reflected in the M-R diagram we mentioned earlier. 
The results are similar when the strength of the interaction is reduced (see also Fig.~\ref{frac-z50-z250-xn50}(f)). 

The results of the impact of the symmetry energy and the self-interaction on the fundamental properties of the neutron star are shown in Fig.~\ref{M-R-eta-all}. From these diagrams, it is clear that the value of the maximum mass depends primarily on the strength of the interaction, while the effect of the symmetry energy is negligible. The symmetry energy appears to play a decisive role in the radius corresponding to the maximum mass, and an even greater role in the radius corresponding to 1.4 solar masses. However, the most dramatic impact of the symmetry energy is seen in the value of the tidal deformability corresponding to these masses, due to the dependence of $\Lambda_{1.4}$ on the fifth power of the corresponding radius. In conclusion, the combination of parameters related to the strength of the interaction and the symmetry energy plays a decisive role in determining the range of tidal deformability values, which can span up to two orders of magnitude. The dependence of the aforementioned quantities on the parameter $\eta$
 becomes increasingly intricate when baryon–dark particle interactions are incorporated. Nonetheless, the contribution of the symmetry energy is non-negligible and must be accounted for in order to attain the highest degree of precision in theoretical predictions.

\section{Concluding Remarks}
The main conclusions of this work can be summarized as follows: (a) Unlike some previous studies, dark neutron decay does not necessarily soften the equation of state. With appropriate tuning of dark sector interactions, the model remains compatible with observed neutron star masses. Although weaker interactions increase the dark particle population and soften the equation of state, the accompanying rise in nuclear symmetry energy counteracts this, leading to a net stiffening. Thus, this decay channel cannot be excluded based on current astrophysical mass constraints,  
(b) The equation of state for mixing neutron star matter and dark matter, which arises from  the dark decay of neutrons, is sensitive to the symmetry energy (for a fixed value of the interaction between the dark particles). The implications of this dependence manifest in the computed values of the stellar radius, and more pronouncedly in those of the tidal deformability, evaluated for a star with a mass of 1.4 solar masses,  
(c) One of the important findings of the present study is that the interaction between baryons and dark particles (when self-interaction is not taken into account) is responsible for the accumulation of a significant fraction of dark particles even at very low densities. This leads to a dramatic alteration of the properties of neutron stars and, most importantly, it can lead to compact objects consisting of a mixture of dark matter and neutron star matter in the mass-gap region, confirming the findings of recent related studies, 
(d)  When dark particles outnumber baryons, new questions arise regarding the particle composition of dense matter, which depends on interactions within the dark sector and with baryons. Since this composition influences key neutron star properties, observational data can impose constraints. How are phenomena such as stellar cooling, the Urca process onset, and the crust-core transition affected? Could additional mechanisms regulate neutron-to-dark particle conversion?
(e) Current researches  investigate  neutron dark  decay in laboratory settings, though the extreme physical conditions within neutron stars  (enormous pressure and density, intense gravitational and magnetic fields, high temperature, e.t.c.) are expected to substantially modify this decay process, 
(f) One possible direction for future research is to employ more advanced equations of state that incorporate the presence for example of hyperons, kaons, and potentially deconfined quark matter. Introducing these additional components would alter the chemical equilibrium conditions, significantly affecting the equation of state and, in turn, the predicted properties of neutron stars.

%

\end{document}